\documentclass{aa}
\usepackage{graphics}
\usepackage{psfig}
\usepackage{rotate}

\def\Tabref#1{Table~\ref{tab:#1}}
\def\VizieR{{\sf VizieR}}

\def\ie{{\em i.e.} }
\def\eg{{\em e.g.} }

\def\secref#1{section~\ref{s#1}}
\def\figref#1{Fig.~\ref{fig:#1}}
\def\tabref#1{Table~\ref{tab:#1}}
\def\HIP{{\em Hipparcos}}

\def\colheader#1{\multicolumn{1}{c}{{\bf #1}}}
\def\ignore#1{}
\def\ifhtx{\iffalse}	% This redefinition is ignored in HTX context.
\begin{document}
\ifhtx
\def\bibcode#1{(#1)}
\tableofcontents
\else	% Define anchors, etc for a printed edition
\def\about{$\sim$}
\def\A#1#2{{#2}\footnote{#1}}
\fi

\def\Aviz#1#2{\A{http://vizier.u-strasbg.fr/#1}{#2}}
\def\Section#1#2{\section{#2}\label{s#1}}
\def\subSection#1#2{\subsection{#2}\label{s#1}}
\def\Avizdup#1#2{ % #2 = 1 to break after parameters
	\Aviz{cgi-bin/VizieR?#1}{a call to \VizieR\ with parameters:
	\ifnum#2>0 \\ \fi
	{\tt #1}}}\def\bibcode#1{}

   \thesaurus{10         % A&A Section 10: Galaxy Dynamics
              (04.01.1 ;
               04.03.01) % Stars: fundamental parameters
               %19.63.1)
        }

\title{The VizieR database of Astronomical Catalogues}

\author{ Fran\c cois Ochsenbein,
  Patricia Bauer,
  James Marcout
  %Fran\c coise Genova\inst{1}
  }

   \institute{CDS,
   Observatoire Astronomique,
   UMR 7550,
   11 rue de l'Universit\'e,
        67000 Strasbourg, France
   %\and
 }
\offprints{F. Ochsenbein {\em (francois}@{\em astro.u-strasbg.fr)}}
\date{December 27, 1999}
\titlerunning{The \VizieR\ database of Astronomical Catalogues}
\authorrunning{Ochsenbein et al.}
\maketitle

\begin{abstract}

\VizieR\  is a  database grouping in an homogeneous
way thousands of astronomical catalogues gathered since dec\-ades by
the Centre de Donn\'ees de Strasbourg (CDS) and participating institutes.
The history and current status of this large collection is briefly presented,
and the way these catalogues are being standardized to fit in the \VizieR\
system is described.
The architecture of the database is then presented, with emphasis
on the management of links and of accesses to very large catalogues.
Several query interfaces are currently available, making use of the
{\em ASU} protocol, for browsing purposes or for use by other data
processing systems such as visualisation tools.

\end{abstract}

%\section{Historical Background}
\section{Introduction}
The { Centre de Donn\'ees astronomiques de Strasbourg} (CDS)
has a very long experience in acquiring,
cross-ident\-ifying, and distributing astronomical data
(Genova et al. \cite{CDS}):
a collaboration for the exchange of what was called
{\em machine-readable} astronomical data started
with the {\em NASA-GSFC} and the {\em Astronomisches Rechen-Institut}
around 1970.
This collaboration has been maintained over this 30 year period,
and collaborations with other institutes for similar exchanges
have been developed.
The volume of data shared of course increased, at a rate
which has been exploding in the recent years.%, tightly related to the
%usage of the Internet.

%{\bf which kind of data}
Compared to the late 60's, where the bulk of the {\em machine-readable}
data consisted in a set of the basic catalogues carefully keypunched,
the situation has changed drastically, now that every instrument or
detector is generating megabytes or gigabytes of daily output.
These huge data sets are hopefully not stored in data centers,
but are processed in the observing center where the expertise
exists to generate the best high-quality archives and
catalogues in a form usable
by astronomers who are not familiar with the instrument.
The Data Centers' role is essentially to collect such
``final'' catalogues, or more generally high-quality data,
\ie data which either were published in the refereed scientific  literature,
or at least a paper describing these data and their context was accepted
for publication in a refereed scientific journal.

Making an efficient usage of the data distributed by the data centers
--- for instance for the analysis of the statistical properties
of some interesting population of stars ---
often requires to {\em combine} data coming from several
data sets;
this operation is far from simple, and this is why the first creation
of CDS was {\sc SIMBAD},
a data-base resulting from the
cross-identification of the major catalogues, later
expanded to thousands of catalogues and to published literature
(see Wenger et al. \cite{simbad}).

The \VizieR\ system results from a different approach: the
astronomical catalogues are kept in their original form,
but homogeneous {\em descriptions} of all these data sets
are provided in order to maximize their usability.
In other words, \VizieR\ relies on an homogenization of the
{\em catalogue descriptions} --- what is also called
{\em metadata}, or data describing other data ---
to transform the set of
{\em machine-readable} astronomical catalogues
into a set of {\em machine-un\-der\-standable} data.
\VizieR\ actually consists in an interface able to
query this set of
{\em machine-un\-der\-standable} astronomical catalogues.

%%%%%%%%%%%%%%%%%%%%%%%%%%%%%%%%%%%%%%%%%%%%%%%%%%%%%%%%%%%%%
\Section{2}{Astronomical Catalogues}
%can now be consulted in \Aviz{cgi-bin/VizieR?-source=}{\VizieR}
% Convert to Copute Means: gawk -f /tmp/a
%#
%BEGIN { FS = "&" }
%{ print; p += $2 ; e += $2*$3 }
%END { printf("\t& %5d\t\& %5.1f\t%% Total\n",p,e/p) }
%%%%%%%%%%%%%%%%%%%%%%%%%%%%%%%%%%%%%%ù

\begin{table*}[hbtp]
%\begin{center}
\begin{tabular}{| l || r r | r r | r r | r r | r r ||r r |} \hline
   Journal & \multicolumn{2}{|c|}{1994}
        & \multicolumn{2}{|c|}{1995}
        & \multicolumn{2}{|c|}{1996}
        & \multicolumn{2}{|c|}{1997}
        & \multicolumn{2}{|c||}{1998}
        & \multicolumn{2}{|c|}{1994--1998}
        %& \multicolumn{2}{|c|}{1999}
        \\
        & Papers & \%El. & Papers & \%El.
        & Papers & \%El. & Papers & \%El.
        & Papers & \%El. & Papers & \%El.
        \\ \hline
   A\&A & 1300  & 1.3   % 1994
        & 1223  & 2.9   % 1995
        & 1394  & 5.3   % 1996
        & 1525  & 6.6   % 1997
        & 1569 	& 4.1   % 1998
	%&{\em 1321} &{\em 3.3} % 1999: 43
	&  5711	&   4.8	% Total
        \\
   A\&AS & 236  & 42.8  % 1994
        & 269  & 42.0   % 1995
        & 438  & 28.4   % 1996
        & 298  & 49.0   % 1997
        & 159  & 43.3   % 1998
	%&{\em 403} &{\em 21.3} % 1999: 86
	&  1164	&  38.9	% Total
        \\
   ApJ(L) & 2064  & 0.3 % 1994
        & 2121  & 0.4   % 1995
        & 2166  & 1.1   % 1996
        & 2255  & 0.8   % 1997
        & 2235  & 0.6   % 1998
	%&{\em 1894} &{\em 0.1} % 1999: 1
	& 10841	&   0.6	% Total
        \\
   ApJS & 255  & 12.9   % 1994
        & 138  & 25.4   % 1995
        & 116  & 22.4   % 1996
        & 115  & 16.5   % 1997
        & 102  & 11.6   % 1998
	%&{\em 99} &{\em 2.0} % 1999: 2
	&   726	&  17.2	% Total
        \\
   PASP & 158  & 7.0    % 1994
        & 161  & 4.3    % 1995
        & 153  & 2.0    % 1996
        & 159  & 2.5    % 1997
        & 181  & 2.8    % 1998
	%&{\em 157} &{\em 157} % 1999: 2
	&   812	&   3.7	% Total
   	\\
   AJ   & 425  & 10.4   % 1994: 44
        & 504  & 14.1   % 1995: 71
        & 477  & 9.2    % 1996: 44
        & 460  & 8.3    % 1997: 38
        & 501  & 9.4    % 1998: 47
	%&{\em 401} &{\em 0.2} % 1999: 1
	&  2367	&  10.3	% Total
        \\
  MNRAS & 656  & 1.4    % 1994: 9
        & 752  & 2.7    % 1995: 20
        & 775  & 0.8    % 1996: 6
        & 833  & 1.6    % 1997: 13
        & 980  & 1.0    % 1998: 10
	%&{\em 831} &{\em 0.5} % 1999: 4
	&  3996	&   1.5	% Total
        \\
\hline
\end{tabular}
\caption{\label{tab:etables}%Existence of electronic tabular data:
Evolution of the annual number of papers, and the percentage of
papers with associated electronic data, for some of the main astronomical
magazines}
%\end{center}
\end{table*}

\begin{table*}[hbtp]
%\begin{center}
%\footnotesize
\scriptsize
\begin{tabular}{|rl|rr|rr|rr|rr|rr|rr||r|} \hline
\multicolumn{2}{|l|}{ {\em Category }}&
        %\multicolumn{2}{|c|}{May 1993}&
        \multicolumn{2}{|c|}{June 1994}&
        \multicolumn{2}{|c|}{June 1995 } &
        %\multicolumn{3}{|c|}{Dec. 1995 }&
        \multicolumn{2}{|c|}{June 1996 } &
        \multicolumn{2}{|c|}{Oct. 1997 } &
        \multicolumn{2}{|c|}{Oct. 1998} &
        \multicolumn{3}{|c|}{Oct. 1999} \\
 & & % $N$ & Mb &
        $N$ & Mb & %\mMb &
        $N$ & Mb & %\mMb&
        $N$ & Mb & %\mMb&
        $N$&Mb %\mMb
        & $N$&Mb %http://vizier.u-strasbg.fr/cgi-bin/VizieR %\mMb
        & $N$&Mb & {\em Std}
        \\ \hline
I  & Astrometric   %&     %Catalogues
        & 151   & 1258  %
        & 158   & 1292  %
        & 167   & 1460  %
        & 199   & 2502  %
        & 207   & 2777  %
        & 210   & 2798  & 113%
        \\
II & Photometric   %&     %Catalogues
        & 144   & 307   %
        & 152   & 320   %
        & 153   & 305   %
        & 187   & 467   %
        & 194   & 525   %
        & 198   & 563   & 110%
        \\
III & Spectroscopic%&     %Catalogues
        & 119   & 162   %
        & 126   & 172   %
        & 125   & 173   %
        & 158   & 233   %
        & 163   & 245   %
        & 170   & 249   & 100%
        \\
IV  & Cross-Identification % &
        & 16    & 89    %
        & 16    & 89    %
        & 15    & 83    %
        & 17    & 91    %
        & 17    & 91    %
        & 17    & 91    & 5%
        \\
V   & Combined Data        %        &
        & 63    & 367   %
        & 63    & 372   %
        & 65    & 365   %
        & 76    & 557   %
        & 84    & 728   %
        & 86    & 842   & 53%
        \\
VI  & Miscellaneous  %   & %Catalogues &
        & 43    & 157   %
        & 49    & 188   %
        & 50    & 502   %
        & 70    & 634   %
        & 71    & 634   %
        & 73    & 653   & 46%
        \\
VII & Non-stellar% & %Objects          &
        & 115   & 361   %
        & 119   & 280   %
        & 122   & 371   %
        & 157   & 425   %
        & 178   & 453   %
        & 180   & 453   & 121%
        \\
VIII & Radio %& %Catalogues            &
        & 24    & 269   %
        & 28    & 269   %
        & 29    & 269   %
        & 39    & 414   %
        & 46    & 615   %
        & 53    & 853   & 51%
        \\
IX   & High-Energy %& %Catalogues&
        & --- & --- %--- &
        & ---  & ---% ---&
        & --- & --- % --- &
        & 6     & 77    %
        & 8     & 79    %
        & 10    & 200   & 10%
        \\
\hline \hline
J/A+A & {\em A\&A}
	& 58    & 2     %
        & 98    & 4     %
        & 158   & 8     %
        & 299   & 16    %
        & 371   & 22    %
        & 424   & 26    & 424%
	\\
J/A+AS          & {\em A\&A Supp.}
        & 123   & 12    %
        & 235   & 24    %
        & 350   & 33    %
        & 544   & 55    %
        & 698   & 73    %
        & 817   & 83    & 817%
	\\
J/AJ            & {\em Astron. J.}
        & 15    & 1     %
        & 91    & 6     %
        & 126   & 10    %
        & 252   & 21    %
        & 295   & 25    %
        & 345   & 31    & 345%
	\\
J/ApJS          & {\em ApJ Suppl.}
        & 13    & 1     %
        & 36    & 4     %
        & 52    & 7     %
        & 111   & 14    %
        & 147   & 16    %
        & 165   & 18    & 165%
	\\
J/(P)AZh & {\em Russian Astron. J.}
        & --- & --- %--- &
        & ---  & ---% ---&
        & --- & --- % --- &
	& 18	& 0.2
	& 26	& 0.5
	& 49	& 0.9	& 49
	\\
\hline
J    & From Journals%&
        & 233   & 17    %
        & 517   & 40    %
        & 766   & 60    %
        & 1404  & 118   %
        & 1771  & 151   %
        & 2087  & 180   & 2087%
        \\
\hline \hline
     & {\em Grand Total} %          &
        & 908   & 2986  %
        & 1228  & 3022  %
        & 1492  & 3588  %
        & 2313  & 5517  %
        & 2739  & 6299  %
        & 3084  & 6882  & 2692%
        \\
\hline
\end{tabular}
\normalsize
\caption{\label{tab:contents}Summary of the evolution of
accessible digital catalogues in the
last five years (number of catalogues and sizes in Mbytes).
The last column gives the number of catalogues with a standardized description
(see \secref{std}).}
%\end{center}
\end{table*}

Jaschek (\cite{cj}) defined a catalogue as
a {\em long} list of {\em ordered data} of a specific kind,
collected for a {\em particular purpose}.
%For the electronic catalogues, the ordering is not really important
%any more, and
What a {\em long} list means has evolved dramatically
in the last decade:
the new way of processing data actually resulted in a tremendous increase
in both the number and the volume of the astronomical catalogues.
%each satellite mission now issues at least one very large catalogue.
To illustrate the evolution in the domain of catalogued {\em surveys},
one can remember that the largest catalogues in the beginning of this century,
called the {\em Durchmusterungen} --- the {\em Bonner},
{\em Cordoba} and {\em Cape} Durchmusterungen ---
provided only a position and a visual estimate of the brightness
for $\sim 1.5\times10^6$ stars,
and required over 50 years to be completed. Today, a catalogue
gathering similar parameters
--- %however
with an accuracy  one order of magnitude better --- is
well represented by the USNO-A2.0 (\cite{usno}) which contains roughly
$5\times10^8$ sources, almost three orders of magnitude larger.
Even larger catalogues are being built: let us quote the
GSC-II (Greene et al., 1998)
which should contain all optical sources brighter than $18^{th}$ magnitude,
which can be estimated to about $2\times10^9$ objects.

The existence of these new {\em mega-catalogues} --- which are, in fact,
rather {\em giga-catalogues} ---
does however not mean that the
old catalogues can just be ignored: virtually any astronomical object
can be subject to variability, maybe over periods of several centuries,
and  the discrepancies between old and newer results have therefore
to be analyzed.

Another important source of %Besides the large catalogues, a large set of
tabular material
%(in terms of reusability)
consists in tables published in the astronomical literature.
These tables are now almost always originally in digital form,
and contain highly processed data which usage can be precious;
access to these electronic data is also essential for maintaining
the large databases like {\sc Simbad} or NED.

The potential interest of the reusability of these tables led
the Editors of the leading astronomical journals to
distribute the tabular material in electronic form.
The first realisations for {\em A\&A} started in 1993
(see Ochsenbein \& Lequeux \cite{aatables}), and \tabref{etables}
summarizes the frequency of the availability of electronic tabular data
among the publications in some of the main astronomical journals
in the recent years: not surprisingly, the {\em Supplement Series},
which were created essentially for the presentation of the observational
results, show a high rate of associated electronic data.

%Evolution with the production of the {\em electronic tables}
%in A\&A, an in other journals.

\Section{2'}{Astronomical Catalogues in the Data Centers}

\subSection{catcontents}{Current Contents}
The growth of the collection of astronomical catalogues
managed by data centers is illustrated
by \Tabref{contents}: the current set of available  catalogues is now around
3,000, with an annual increase about 15\%.
Note that the entity designated as a ``catalogue''
can represent a table of about 100 entries
(\eg the list of galactic globular clusters),
as well as a multimillion source catalogue (\eg the USNO-A2.0).

In \Tabref{contents}, the catalogues are grouped according
to {\em categories} which were defined in the 70's,
when the bulk of astronomical
studies were dealing with the properties of stars in the optical
wavelength domain. Rather than defining regularly a new classification scheme
following the evolution of the discipline, it was decided,
in agreement with the other data centers,
to assign designations to electronic tables
according to the published paper, and to reserve the assignment
in the ``traditional'' categories
to somewhat important catalogues or compilations.
Simultaneously, it was decided to assign {\em keywords} to each
catalogue, in order to allow easy retrieval of catalogues with similar
contents and purposes.

Note that, if most of the catalogues contain data related to the observation of
astronomical sources,
%either actual observation or compilationcatalogues,
%--- like the resulting catalogue of the {\em Hipparcos} mission --- or from a
%compilation.
other types of data are also available, generally grouped in
the `Miscellaneous'' (VI) category:
catalogues of atomic data like wavelength tables or
results of the {\em Opacity Project}, tabulated results
of stellar evolution models, ephemeris elements, etc\dots

\subSection{catusage}{Usage of astronomical catalogues}

\begin{table}[htbp]
%\begin{tabular}{|l| rr r r|}
\begin{tabular}{|l| rr r |}
\hline
Period & Files & Gbytes & Nodes \ignore{& Load }\\ \hline
1993Jan--1993Dec & 6106 & 1.5 & 458	\ignore{& 0.1 Kb/s }\\
1994Jan--1994Dec & 23696& 6.1 & 1599	\ignore{& 0.2 Kb/s}\\
1995Jan--1995Dec & 57314  & 11.4 & 4022 \ignore{& 0.4 Kb/s}\\
%1995Jun--1996May &  70100 & 15.9 & 4928 \ignore{& 0.5 Kb/s }\\
1996Jan--1996Dec & 71300  & 19.8 & 4953 \ignore{& 0.6 Kb/s }\\
%1996Jun--1997May &  90000 & 21.4 & 5464 \\
1996Oct--1997Sep & 143000 & 43.5 & 6279 \ignore{& 1.4 Kb/s}\\
%1997Jan--1997Dec & 134618 & 49.5  & 4959 \ignore{& 2.4 Kb/s}\\
1997Oct--1998Sep & 308840 & 74.5 & 9780 \ignore{& 2.4 Kb/s}\\
%1998Jan--1998Dec & 384465 & 69.3  & 10360 \ignore{& 2.4 Kb/s}\\
1998Oct--1999Sep & 538407 & 77.1 & 10146 \ignore{& 2.5 Kb/s}\\
\hline
\end{tabular}
\caption{\label{tab:ftp} Yearly traffic on the CDS catalogue ftp server
{\em(internal and mirror traffic excluded)}}
\end{table}

One of the main goals of the CDS is to promote the usage of the reliable
astronomical catalogues to the astronomical community.
The ``Catalogue Service'' has been one of the major CDS services since
the beginning of the CDS activity, and used to distribute catalogues
on magnetic tapes and floppies; the service has been implemented
on the network as a FTP server in March 1992, generating
immediately a large increase in
the number of distributed files. The FTP activity is still increasing at a high
rate, as can be inferred from \Tabref{ftp}: the current traffic
is equivalent to a copy of the whole collection every month.

It is also interesting to quote those catalogues which are the
most frequently copied from the CDS archives, %which are
summarized in \Tabref{top} for the last two years: not surprisingly,
{\em surveys}, and what Jaschek (\cite{cj}), in his section 5.2,
designates as {\em General Compilation Catalogues},
are among the most popular catalogues.
It is also interesting to note the large number of copies
of the GSC catalogue (about 300 Mbytes): it was copied by over 500 nodes
in the last 12 months, which is 4 times more than in the previous year;
this could indicate that
catalogues of this size can be quite easily managed on
small computers nowadays.

\begin{table}[htbp]
\scriptsize
\begin{tabular}{| r r p{0.34\textwidth}|} \hline
 \multicolumn{2}{|c}{Number of Nodes} & {Catalogue designation and short title} \\
 1999&(1998)& \\ \hline
  879 &(750)&(I/239)  Hipparcos \& Tycho Catalogues \\
  502 &(123)&(I/220) The HST Guide Star Catalog, V1.1 (Lasker+ 1992) \\
  293&(165)& (VI/87) Planetary Ephemerides (Chapront+ 1996) \\
  284&(241)&(I/131A)  SAO Star Catalog J2000 (SAO Staff 1966; USNO, ADC 1990) \\
  248 &(60)&(I/197)  Tycho Input Catalogue, Revised version (Egret+ 1992) \\
  203&(221)&(VII/118)  NGC 2000.0 \\
  195&(162)&(V/50)  Bright Star Catalogue, 5th Revised Ed. (Hoffleit+, 1991)\\
  173&(145)&(VI/80) Opacities from the Opacity Project    (Seaton+, 1995)\\
  169&(134)&(I/246) The ACT Reference Catalog (Urban+ 1997) \\
  126& (142)&(V/70A)    Nearby Stars, Preliminary 3rd Version  (Gliese+ 1991) \\
  124 &(120)&(VI/81) Planetary Solutions VSOP87 (Bretagnon+, 1988) \\
  112 &(73)&(VII/207) Quasars and Active Galactic Nuclei (8th Ed.)
  	(Veron+ 1998)\\
  102& (134)&(II/214A) Combined General Catalogue of
	Variable Stars (Kholopov+ 1998) \\
  101 &(76)& (VII/155) Third Reference Cat. of Bright Galaxies (RC3)
  	(de Vaucouleurs+ 1991) \\
  100 &(75)& (VI/79)  Lunar Solution ELP 2000-82B (Chapront-Touze+, 1988) \\
  99&(153)&(VI/69)   Atomic Spectral Line List    (Hirata+ 1995)\\
  97&(149)&(V/95)  SKY2000 - Master Star Catalog (Myers+ 1997) \\
  90&(118)&(I/196) Hipparcos Input Catalogue, Version 2   (Turon+ 1993) \\
  %(98)&(I/146)     PPM North Star Catalogue              (Roeser+, 1988)\\
  %(95)&(V/84)  Strasbourg-ESO Catalogue of Galactic Planetary Nebulae
  	%(Acker+, 1992)\\
  %  (91)&(I/237) The Washington Visual Double Star Catalog, 1996.0
  	%(Worley+, 1996)\\
  %(85)&(II/183)  UBVRI Photometric Standards (Landolt 1992) \\
  %(84)&(I/245) Orbital Elements of Minor Planets 1998 (Batrakov+ 1997)\\
\hline\end{tabular}
\normalsize
\caption{\label{tab:top} Catalogues which have been the most frequently
copied}
\end{table}

%%%%%%%%%%%%%%%%%%%%%%%%%%%%%%%%%%%%%%%%%%%%%%%5
\Section{std}{Standardized Description of Astronomical Catalogues}

\begin{figure*}[hbtp]
\begin{center}
\scriptsize
\begin{verbatim}
I/221               The Magellanic Catalogue of Stars - MACS    (Tucholke+ 1996)
================================================================================
The Magellanic Catalogue of Stars - MACS
     Tucholke H.-J., de Boer K.S., Seitter W.C.
    <Astron. Astrophys. Suppl. Ser., 119, 91-98 (1996)>
    <The Messenger 81, 20 (1995)>
    =1996A&AS..119...91T =1995Msngr..81...20D
================================================================================
ADC_Keywords: Magellanic Clouds ; Positional data

Description:
    The Magellanic Catalogue of Stars (MACS) is based on scans of ESO
    Schmidt plates and contains about 244,000 stars covering large areas
    around the LMC and the SMC. The limiting magnitude is B<16.5m and the
    positional accuracy is better than 0.5" for 99% of the stars. The
    stars of this catalogue were screened interactively to ascertain that
    they are undisturbed by close neighbours.


File Summary:
--------------------------------------------------------------------------------
 FileName    Lrecl    Records    Explanations
--------------------------------------------------------------------------------
ReadMe          80          .    This file
lmc             52     175779    The Large Magellanic Cloud
smc             52      67782    The Small Magellanic Cloud
--------------------------------------------------------------------------------

Byte-by-byte Description of file: lmc smc
--------------------------------------------------------------------------------
   Bytes Format  Units   Label    Explanations
--------------------------------------------------------------------------------
   1- 12  A12    ---     MACS     Designation
  14- 15  I2     h       RAh      Right Ascension J2000 , Epoch 1989.0 (hours)
  17- 18  I2     min     RAm      Right Ascension J2000 (minutes)
  20- 25  F6.3   s       RAs      Right Ascension J2000 (seconds)
      27  A1     ---     DE-      Declination J2000 (sign)
  28- 29  I2     deg     DEd      Declination J2000 , Epoch 1989.0 (degrees)
  31- 32  I2     arcmin  DEm      Declination J2000 (minutes)
  34- 38  F5.2   arcsec  DEs      Declination J2000 (seconds)
      40  I1     ---     Npos     Number of positions used
  42- 46  F5.2   mag     Mag      []?=99.00 Instrumental Magnitude
                                        (to be used only in a relative sense)
      48  I1     ---     PosFlag  [0/1] Position Flag   (0: ok,
                                        1: internal error larger than 0.5")
      50  I1     ---     MagFlag  [0/1] Magnitude Flag  (0: ok,
                                        1: bad photometry or possible variable)
      52  I1     ---  BochumFlag *[0] Bochum Flag
--------------------------------------------------------------------------------
Note on BochumFlag: 1 if in Bochum catalog of astrophysical information
    on bright LMC stars (yet empty)
--------------------------------------------------------------------------------

Author's address:
    Hans-Joachim Tucholke    <tucholke@astro.uni-bonn.de>

================================================================================
(End)            Hans-Joachim Tucholke [Univ. Bonn]                  20-Nov-1995
\end{verbatim}
\normalsize
\caption{\label{fig:readme}Example of a documentation {\tt ReadMe} file }
\end{center}
\end{figure*}

Making use of the data contained in a set of rapidly evolving catalogues,
as illustrated by \Tabref{contents},
raises the problem of accessing and {\em understanding} accurately
the parameters contained in catalogues which are constantly improved.
%have nowadays rather short lifetimes.
Typical questions to be addressed are:
does the catalogue contain colours; if yes what is their reliability;
are they expressed in a well-known standard system; are they taken from
other publications or catalogues;
how can the associated data file be processed?
All these details which  describe the data --- the {\em metadata} ---
are traditionally
presented in the introduction of the printed catalogue, or
detailed in one or several published papers presenting
and/or analyzing the catalogued data.

Metadata play therefore a fundamental role: first the
scientists have to get information about the
{\em environment} of the data in order to make
their judgement about the suitability of the data for their project, such as:
date and/or method of acquisition, related publications,
estimation of the internal and external errors, %possible distorsions,
purpose of the data collection, etc.;
but also a minimal knowledge of the metadata is required by
the data processing system
%to understand the data not only as arrays of numbers
in order to merge or compare data from different origins ---
for instance, the comparison of data expressed in different units
requires a unit-to-unit conversion which can be performed
automatically only if the units are specified unambiguously.

This need for a description which is readable both by a computer
and by a scientist led to a standardized way of documenting astronomical
catalogues and tables,
promoted by CDS from 1993
in the form of a dedicated {\tt ReadMe} file associated to each catalogue
(Ochsenbein \cite{readme}).
An example of such a file is presented in \figref{readme}:
it is a plain ascii file, quite easy to interpret for a scientist,
and at the same time structured enough to be interpreted
by a dedicated software.
The {\tt ReadMe} description file
starts with a {\em header} specifying the basic references
--- title, authors, references ---
and contains a few key sections introduced by standard titles like
{\tt Description:} or {\tt Byte-by-byte Description of file:}.
Such a file is relatively easy to produce by someone who
knows the catalogue contents.
The example of
\figref{readme} represents the documentation of
a very simple catalogue, made of just two data tables, each
with a small  set of parameters.
The \Aviz{cgi-bin/Cat?I/239}{output catalogue of the \HIP\ mission}
is an example of a much more complex catalogue: it
is composed of two fundamental large tables (HIP with $10^5$ stars
and TYC with $10^6$ stars) and includes
a dozen of annex tables, but can still be described by the
the same kind of simple standardized documentation.

The most important part of the {\tt ReadMe} file
is the \\{\tt Byte-by-byte Description} which details the table structures
in terms of {\em formats}, {\em units}, column naming or {\em labels},
{\em existence of data}
(possibility of unspecified or {\em null} values), and brief explanations.
Among the conventions, some fundamental parameters are assigned
fixed labels like sky coordinates (components of right ascension {\tt RA}...
and declination {\tt DE}... in \figref{readme}); a {\em prefix} convention,
detailed in \tabref{prefix}, is also used to specify obvious relations
between a value, its mean error, its origin, etc...

\begin{table}[hbtp]
\begin{center}
\scriptsize
%\begin{tabular}{lll} \hline
\begin{tabular}{lp{.36\textwidth}} \hline
\colheader{Symbol} & \colheader{Explanation} %& \colheader{Default Limits}
	\\ \hline
%{\tt 2\_{\em label}} & $\chi^2$ value
%		on parameter {\em label} \ignore{& $\geq0$} \\
{\tt a\_{\em label}} & {\em aperture} used for
		parameter {\em label} \ignore{& $\geq0$} \\
{\tt E\_{\em label}} & mean error (upper limit)
		on parameter {\em label} \ignore{& $\geq0$} \\
{\tt e\_{\em label}} & mean error ($\sigma$)
		on parameter {\em label} \ignore{& $\geq0$} \\
{\tt f\_{\em label}} & {\em flag}
		on parameter {\em label} \ignore{& } \\
{\tt l\_{\em label}} & {\em limit flag}
		on parameter {\em label} \ignore{& {\tt[<>]}} \\
{\tt m\_{\em label}} & {\em multiplicity index}
		on parameter {\em label} to resolve ambiguities \ignore{& } \\
{\tt n\_{\em label}} & {\em note} (remark)
		on parameter {\em label} \ignore{& } \\
{\tt o\_{\em label}} & number of {\em observations}
		on parameter {\em label} \ignore{& $\geq0$} \\
{\tt q\_{\em label}} & {\em quality}
		on parameter {\em label} \ignore{& } \\
{\tt r\_{\em label}} & reference (source) for parameter {\em label}
		\ignore{& } \\
{\tt u\_{\em label}} & {\em uncertainty flag}
		on parameter {\em label} \ignore{& {\tt[ :]}} \\
{\tt w\_{\em label}} & {\em weight}  of parameter {\em label}
		\ignore{& $\geq0$} \\
{\tt x\_{\em label}} & unit in which parameter {\em label} is
	expressed \ignore{&  } \\
\hline\end{tabular}
\normalsize
\caption{\label{tab:prefix}Conventions used for {    label} {\em prefixes} }
\end{center}
\end{table}

This standardized way of presenting the metadata
pro\-ved to be extremely useful, especially for {\em data checking}
and {\em format conversion}:
many errors were detected in old catalogues simply because a
general checking mechanism became available.
Tools have been developed  for generating a Fortran source
code which loads the data into memory,
or for converting the data into the FITS format
which is presently the most ``universal'' data format understood by
data processing systems in astronomy --- but unfortunately a data
format which is not convenient outside this context
(see \eg Gr\o sb\o l et al. \cite{fits}).

During the six years since this standardized way of describing
astronomical catalogue has been defined,
over 2,600 astronomical catalogues have been described
by means of this {\tt ReadMe} file, and the same conventions have been adopted
by the other astronomical data centers and journals for
the electronic publication of tables.
The present (October 1999) figures of the amount of standardized
catalogues are summarized in the rightmost column
of \Tabref{contents}; previous figures
were presented in an earlier paper (Ochsenbein \cite{russie}).

It is expected, in the future,
that the authors will supply the documentation of their data
in this simple form;
it is already the case for a very significant fraction of
the tables mailed to the CDS,
and in order to help the authors,
template files as well as a few \A{http://vizier.u-strasbg.fr/doc/submit.htx}
{tips on how to create the {\tt ReadMe} file are accessible on the Web}.
The {\tt ReadMe} files and the data files are then checked by a specialist,
who contacts the authors if errors are detected or when changes are
necessary to increase the clarity or homogeneity of the description.

%%%%%%%%%%%%%%%%%%%%%%%%%%%%%%%%%%%%%%%%%%%%%%%%%%%%%%%%%%%%%
\Section{4}{\VizieR\ Organisation}

\A{http://vizier.u-strasbg.fr/}{\VizieR}  is a natural extension
of the usage of the metadata stored in  the
{\em ReadMe} files, as an implementation of these metadata
in terms of tables managed by a relational database management
system (RDBMS).

The first prototype of \VizieR\ was the result of a fruitful collaboration
between {\em ESIS} (European Space Information System,
a project managed by ESRIN, a department of the European Space Agency)
and the CDS; \VizieR\ has been under full responsibility of CDS since
January 1996.
It was presented at the 1996 AAS meeting (Ochsenbein et al., \cite{aas96}),
and became fully operational in February 1996.
This prototype has been significantly upgraded in May 1997, just
in time for the implementation of the final catalogues of the Hipparcos
mission. %, described in somewhat more
%details at an ADASS conference (Ochsenbein \cite{adass97}).
%The described version of the {\em META} dictionary was first
%implemented in 1997 (Ochsenbein \cite{adass97}).
%\VizieR\ was first presented at the AAS meeting
%(Ochsenbein et al., \cite{aas96})
%in the beginning of 1996, and is described in somewhat more
%details at an ADASS conference (Ochsenbein \cite{adass97}).
The number of catalogues
accessible within the \VizieR\ system has grown since that time
to 2,374 catalogues (\Tabref{vizcounts}).

%\Section{5}{\VizieR\ Metadata}

The core of \VizieR\ consists in the organisation
of the {\em meta dictionary}, i.e. the set of metadata extracted from the
standardized {\tt ReadMe} descriptions discussed in \secref{std}.
There are however two main problems which had
to be solved: the access to very large catalogues (larger than a
few million rows) for
which RDBMS proved to be  inefficient, requiring therefore
dedicated search methods, % briefly presented in \secref{L};
and the generation of {\em links} allowing to connect two related pieces of
information, like other tables in the same catalog, or
spectra, images from remote services, etc.%, presented in \secref{6}.

\subSection{meta}{META dictionary}

\begin{table}[htb]
\begin{tabular}{|ll|r|r|}\hline
\multicolumn{2}{|l|}{\VizieR\ contents}	& All 	& Dealing with objects \\
\multicolumn{2}{|l|}{in terms of:}	& Catalogues	& having positions\\ 				\hline
\multicolumn{2}{|l|}{{\em Catalogues:}}	& 2374	& 1247 \\
\multicolumn{2}{|l|}{{\em Tables:}}	& 6071	& 1929 \\
\multicolumn{2}{|l|}{{\em Columns:}}	& 77260	& 30261 \\
\multicolumn{2}{|l|}{{\em Rows:}}	& $1.17\times10^9$ &$1.16\times10^9$\\
	& {\em(without megacatalogs)} 	& $40.3\times10^6$ &$31.6\times10^6$\\
\hline\end{tabular}
\caption{\label{tab:vizcounts}Summary of the \VizieR\ contents (November 1999)}
\end{table}

The meta-dictionary consists in 3 main tables
detailed below, and about 20 annex tables, all stored in a relational
database:

\begin{enumerate}
\item	{\em METAcat} describes the {\em catalogues}, a
	{\em catalogue} being defined as a set of related tables
	published together: typically a catalogue
	gathers a table of observations,
		a table of mean values, a table of references,
		a list of related images, etc\dots;
	{\em METAcat} details the authors, reference,
	title, explanations of each stored catalogues.
	This table contains currently 2,374 rows (\Tabref{vizcounts}).
	%of the \about\ 2400 stored catalogues.
\item	{\em METAtab} describes each {\em data table} stored in \VizieR:
        table caption, number of rows, how to
	access the actual data, the equinox and epoch of the coordinates,
	etc\dots
	This table contains currently 6,071 rows (\Tabref{vizcounts}) ---
	\ie the average catalogue is made of 2.6 tables.
\item	{\em METAcol} details each
	of the  77,260 columns (\Tabref{vizcounts}) currently
	stored in \VizieR: column name or {\em label},
	the textual {\em explanation} of the column contents,
	{\em datatypes} (numeric or character) and storage mode
	within the database (integer or floating-point, maximal length of
	strings, etc), {\em units} in which the data are stored in the
	data-base and {\em units} in which the data are presented
	to the user, edition formats, and a few flags used for searches (e.g.
	column used as primary key) or data presentation (e.g. column
	to be displayed in the default presentation of the result).
	The average table is therefore made of $\sim 12.7$ columns ---
	in fact $\sim 11.7$ because each table contains an {\em identification}
	column in addition to the original set of columns.
\end{enumerate}
Note that, since the set of {\em META} tables is itself described in \VizieR,
the meta-dictionary can be viewed and queried like any
of the catalogues stored in \VizieR\ --- allowing to locate easily \eg
tables with a large number of rows, or catalogues having the words
{\em mass loss} in the description of one of their columns,
etc\dots

The annex tables of the meta-dictionary contain some definitions,
like the list of known data-types {\em(METAtypes)} and
keywords {\em(METAkwdef)};
or other details like the acronyms
used to designate well-known catalogues like {\em HIP}, {\em GSC}
\dots {\em(METAcro)},
the keywords associated to each catalogue {\em(METAkwd)}, detailed notes
and remarks {\em(METAnot)}, or the list of those objects which are
individually quoted in the {\tt ReadMe} files {\em(METAobj)}.
A special indexing scheme {\em(METAcell)}, explained briefly in
\secref{cell}, was built to locate the existing objects in all catalogues
in a single run.
Details on how to generate links are stored in the {\em METAmor} table.

\subSection{link}{Links in \VizieR}
The interest of having a {\em link}, or an {\em anchor} in HTML terms,
becomes obvious when a table contains a column representing
a reference to an original paper, % quoting the result detailed in another table,
as for example in \Aviz{cgi-bin/VizieR?-source=7207/table1}{V\'eron
and V\'eron's
compilation of quasars}: once the rules to transform the contents
of this column into an actual link to \eg the
\A{http://adswww.harvard.edu/}{ADS bibliographic service} is set up,
details about the authors and references,
or even the full article, can then be displayed on the screen by a
simple mouse click. Another frequent example is the possible expansion
of some footnote symbol into the
lengthy note detailed in some other table.

The links existing in \VizieR\ may be classified in the following categories:
\begin{enumerate}
\item	{\em hard-wired links} which are part of the standard description
	presented in \secref{std}, like the existence of
	notes (stored in the {\em METAnot} table), or the {\tt r\_} prefix
	(\tabref{prefix}) %in a column label
	which indicates a reference
	which may be detailed in  a table of references;
\item	{\em internal links} which connect tables of the same catalogue:
	such links may be expressed in terms of {\em keys} in the
	RDBMS terminology (definitions of columns as primary and/or foreign keys),
	by the existence of {\em note flags}, or
	by more complex relations stored in the {\em METAmor} table.
	Another type of {\em internal link} allows one to retrieve
	the spectra or images which are part of the catalogue,
	but which are stored as separate files.
\item	{\em \VizieR\ links} which refer to another catalogue
	within the \VizieR\ system;
\item	{\em external links} which refer to any other service, like
	bibliographic services, external databases or archives,
	image servers, etc.
\end{enumerate}

While links of the first 3 categories can easily be maintained,
the maintenance of the {\em external links} depends on modifications
which are completely outside \VizieR's control.
These external links are maintained by the {\em GLU} system
(Fernique et al., \cite{GLU}), a system which %basically
(i) allows one to use {\em symbolic names}  instead of hard-coded URLs,
and (ii) translates these symbolic names
with the help of a {\em distributed} dictionary
in which the service providers keep up the descriptions of their
own services  only
in terms of URL addresses and actual presentation of the query
parameters.

\subSection{pipeline}{\VizieR\ feeding pipeline}
On the average, about one new catalog -- or 2.6 tables -- is added
{\em daily} into VizieR. Such figures imposed the
following constraints on the addition of new tables into \VizieR:
\begin{enumerate}
\item	no human intervention is required to populate the database
	(the meta dictionary and the data tables):
	all meta-data related
	to a catalogue can be found or computed on the basis of
	documentation and configuration files which are read by
	the \VizieR\ feeding pipe-line ;
\item	we rely as much as possible on the
	{\em standardized description} of the catalogues presented
	in \secref{std}:
	this means that the configuration file associated to
	each catalogue should be minimized, i.e.
	as few {\em ad-hoc} details as  possible
	should be needed besides the {\tt ReadMe} files.
%\item	the system must be {\em efficient}
%	and guarantee that the frequently asked searches
%	(typically from a position on the sky, but \VizieR\
%	also includes frequently used reference tables of atomic
%	radiations for which
%	the typical search is based on the wavelength)
%	can be achieved in less than a second,
%	regardless of the catalogue size.
\end{enumerate}

The actual delay required to ingest a new catalogue
into the system is currently estimated to
something between a few minutes and several days for the preparation of
	the {\tt ReadMe} description file, depending on the
	initial presentation supplied by the authors and on the
	catalogue complexity --- the delay can
	be occasionally longer when problems are encountered, requiring
	interactions with the authors;
and	a few seconds up to an hour for the actual ingestion into \VizieR\
	from the standardized files.

%\Section{L}{Access to Very Large Catalogues}
\subSection{L}{Access to Very Large Catalogues}
\begin{table}[htb]
\begin{tabular}{lrp{0.28\textwidth}} \hline
Acronym & Rows & Catalogue designation \\
      & ($\times10^6$) \\ \hline
USNO-A1.0 & 488.0 & The USNO-A1.0 Catalog (Monet 1997) \\
USNO-A2.0 & 526.3 & The USNO-A2.0 Catalog (Monet 1998),
			calibrated against Tycho data \\
GSC1.1 & 25.2 & HST Guide Star Catalog, 1992 version \\
GSC1.2 & 25.2 & HST Guide Star Catalog, 1996 version \\
GSC-ACT& 25.2 & HST Guide Star Catalog, calibrated against Tycho data${}^\dag$\\
2MASS & 20.2 & $2\mu m$ All Sky Survey, Spring 1999 release
		(Skrutskie et al., \cite{2mass}) \\
DENIS & 17.5 & Deep Near-IR Survey first release
		(Epchtein et al., \cite{DENIS})\\
\hline\end{tabular}
{\footnotesize $\dag$ calibration made by the {\em Pluto project} \\
	(http://www.projectpluto.com/gsc\_act.htm)}
\caption{\label{tab:megacat} Large catalogues currently
implemented in \VizieR}
\end{table}

The second challenge is to open a fast  access for querying
the {\em mega-catalogues}
introduced in \secref{2}. This denomination was
somewhat arbitrarily assigned to catalogues having $10^7$ or more
rows. Such large catalogues are essentially surveys used as
{\em reference catalogues},
typically to find all objects detected in some region of the
sky under some conditions of wavelength, time, object structure, etc.
The set of such catalogues currently implemented is summarized
in \Tabref{megacat}, but this set will grow rapidly in the near future
with the continuation of the infra-red surveys,
and the emergence of surveys presently in preparation
(SLOAN, GSC-II, NVSS, \dots).

The limit of $10^7$ rows corresponds to a limit in performance and
time required to ingest the tables into
the relational data\-ba\-ses;
the largest table, in terms of number of rows, currently stored in
\VizieR\ is the {\em AC2000} catalog (Urban et al., \cite{ac2000}),
with $4.62\times10^6$ rows.

The method used to access these very large catalogues %basically
consists in grouping the objects within carefully designed {groups}
based essentially on the location in the sky,
followed by a lossless compression obtained by replacing the
actual values by offsets within the group;
details about the actual results and performances are
described in another paper (Derriere \& Ochsenbein,
\cite{adass99-poster}).
Each very large catalogue has presently its own organisation
which depends on its actual column contents,
and therefore requires a dedicated program for accessing it.
\VizieR\ stores in its {\em META} dictionary (see \secref{meta})
which program has to be called to actually access the catalogue,
and the description of the columns as they are returned from
the dedicated program.

\subSection{cell}{Accessing all catalogues from a position in the sky}
In order to allow a fast answer to the question:
{\em find out all objects for all available catalogues around
some target position}, an indexing mechanism is necessary.
The total number of object positions currently stored in \VizieR,
excluding the {\em megacatalogues}, is
about $32\times10^6$ (\Tabref{vizcounts}); a classical indexation, in
terms of relational DMBS, shows very poor performances
especially in the updating phase: the addition of a new catalogue
can require up to 4.6 millions modifications or additions --
which becomes dramatically slow.

The method adopted for this indexation consists first in a mapping
of the celestial coordinates into a set of boxes using a
hierarchical spherical-cubic projection similar to
the techique used by {\sc Simbad}
(Wenger et al., \cite{simbad}), but down to a level 8 which corresponds
to a granularity of about $20'$, or
$6\times4^8$ ($\simeq 4\times10^5$) individual boxes.
The list of catalogues which exhibit sources in the region of the sky
covered by the box is then stored for each of the defined boxes,
allowing therefore a fast answer to the question:
``what is the list of catalogues
which have a fair chance of having at least one source close to
a specified target~?'' The final step consists in looking successively
into the matching catalogues.

The method offers the particularity of being {\em hierarchical}:
6 boxes are defined at level 0, 24 at level 1, \dots,
and going down one step in the
hierarchy consists in dividing each box into four parts.
The indexing mechanism recursively groups contiguous non-empty boxes
represented by a single box
at the upper level, meaning that a dense survey covering the whole
sky is just represented by the 6 boxes of level 0 in this index.
In practice, the 1247 catalogues with positions are summarized in
this index by $3.9\times10^6$ elements (to be compared to the
$31.6\times10^6$ sources in \Tabref{vizcounts}), \ie an average of
3,000 elements per catalogue.

\subSection{vizcontents}{Current Contents}

\begin{figure}[htb]
%\resizebox{\hsize}{!}{\includegraphics{vizh1.ps}}
\begin{center}
\psfig{figure=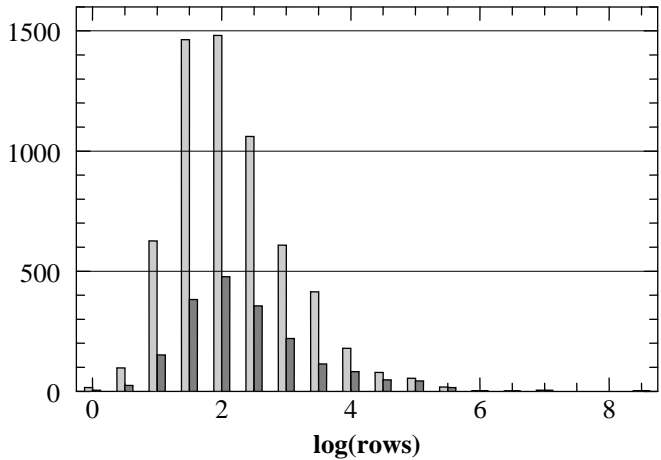,width=\hsize}
\end{center}
%\psfig{figure=viz-histo.eps,width=8cm,angle=90}
%\resizebox{\hsize}{!}{\rotatebox{+90}{\includegraphics{viz-histo.eps}}}
%\vspace*{-3cm}
%\resizebox{\hsize}{!}{\rotatebox{-90}{\includegraphics{viz-histo.eps}}}
\caption{\label{fig:vizhisto}Histogram of the number of rows among
  \VizieR\ tables (the darker bars correspond to tables containing celestial coordinates).}
\end{figure}

The status of \VizieR\ contents is presented in
\Tabref{vizcounts}, where we distinguished those tables representing
data about actual {astronomical objects} which can be accessed by
their position in the sky. In terms of number of available records,
those containing celestial positions represent over 78\% even
when the {\em megacatalogs} are omitted, even though only 32\% of the
tables are concerned. In other words, the average table dealing with
actual astronomical objects contains around 16,000 rows ---
a theoretical mean, as can be seen from the histogram of the table
populations in \VizieR\ represented in \figref{vizhisto} which shows
a modal value around tables of 100 objects.

%, represented
%by 6,071 tables, and a total of 77,260 described columns (November 1999)
%(\Tabref{vizcounts}).
\Section{U}{\VizieR\ Interfaces}

\begin{figure}[htb]
\resizebox{\hsize}{!}{\includegraphics{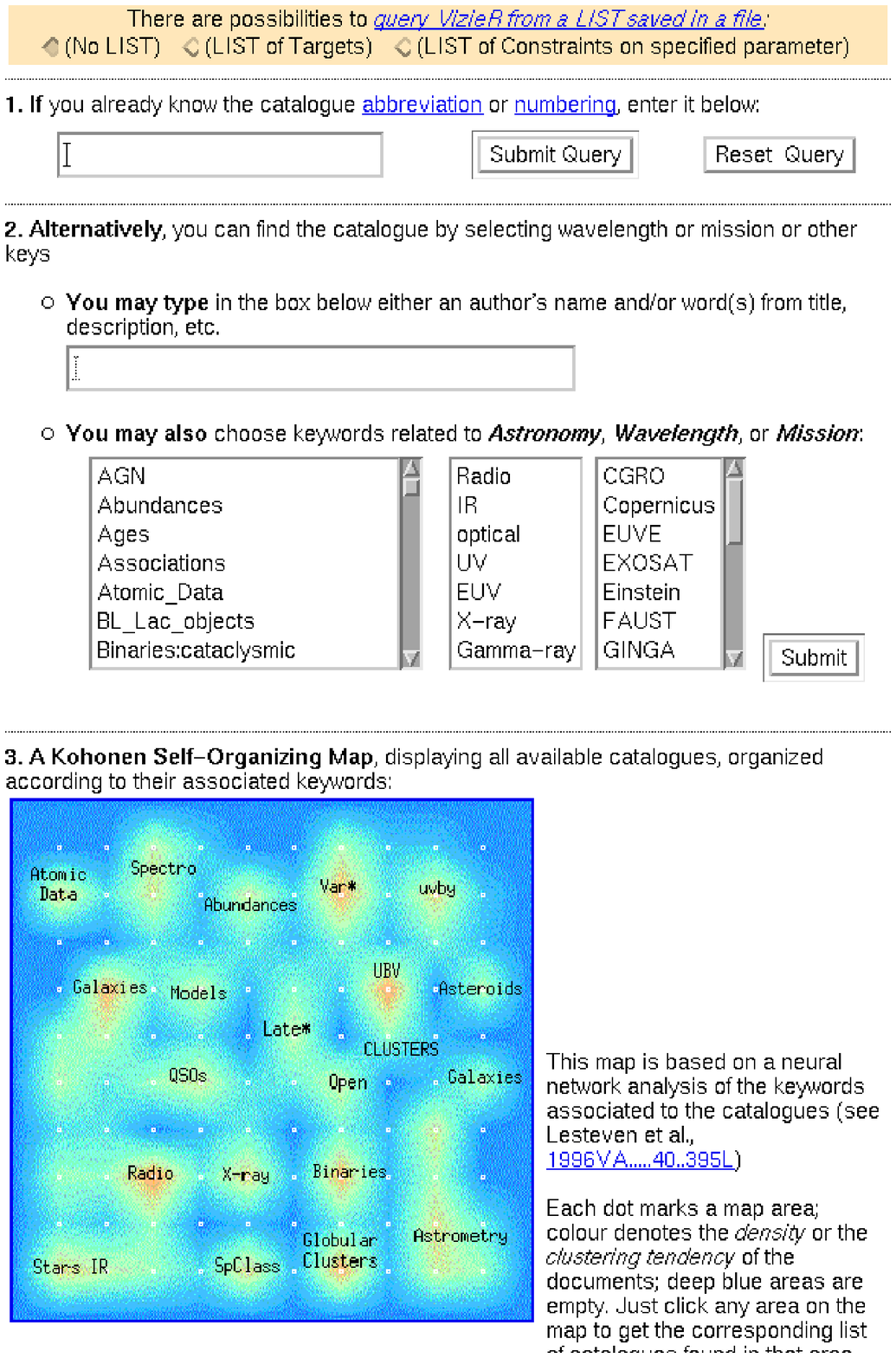}}
%\resizebox{\hsize}{!}{\includegraphics{viz-fig1.ps}}
\caption{\label{fig:netscape}Excerpt of the \VizieR\ first search page
  }
\end{figure}

Several interfaces are currently available for an access to the
data stored in \VizieR: directly from a Web browser,
via a construction of the query using the {\em ASU} conventions,
or the developing {\em XML } interfaces.

\subSection{netscape}{Access from a Browser}
From a WWW-browser,  a ``standard query'' in \VizieR\ consists in a few steps:
\begin{enumerate}
\item	
	Locate the interesting catalogues in the
	\Aviz{cgi-bin/VizieR}{\VizieR\ Service}.
	This can be done in various ways illustrated in \figref{netscape}:
	from well-known catalogue
	acronyms like {\em HIP} or {\em GSC},
	from a choice in the set of predefined keywords,
	from authors' names, or from a self-organizing (or
	Kohonen) map
	constructed on the basis of the keywords attached to
	the catalogues (Poin\c cot et al. \cite{kohonen}).
	New possibilities for locating catalogues of interest for the user
	are currently under development.% (see \secref{7}).

\item	Once a catalog table -- or a small set of catalog tables ---
	is located (for instance the
	\Aviz{cgi-bin/VizieR?-source=I/239/hip\_main}{\HIP\ Catalog}
	resulting from the \HIP\ mission),
	{\em constraints} about what to search and how to
	present the results can be specified, as:
	\begin{itemize}
	\item	constraints based on the celestial coordinates, i.e.
		location in the neighbourhood of a target specified
		by its actual coordinates in the sky, or by
		one of its name as known in {\sc Simbad} (see Wenger et al.,
		\cite{simbad})
	\item	any other constraint on any of the columns
		of the table(s), like a minimal flux value,
		or the actual existence of some parameter
		(non-{\em NULL}  value)
	\item	which columns are to be displayed, and in which order
		the matching rows  are to be presented.
	\end{itemize}
	By pushing the appropriate buttons, it is for instance
	easy to get the
	\Aviz{cgi-bin/VizieR?-source=I/239/hip\_main\&-sort=-\-Plx\&Plx=\%3e=200}
	{list of \HIP\ stars closer than 5 parsecs to the Sun,
	ordered by their increasing distance}.

\item	Obtaining full details about one row is achieved by a %simple
	mouse click in the first column of the result: for instance,
	the first row of the search for nearby stars described above
	leads to the
	\Aviz{cgi-bin/VizieR-5?-source=I/239/hip\_main\&HIP=70890}
	     {\VizieR\ Detailed Page with \HIP\ parameters and
	     their explanations concerning
	     Proxima Centauri}.

\item	Finally, there may be correlated data, like notes or remarks,
	references, etc\dots. In our example, Proxima Centauri
	is related to the $\alpha$ Cen multiple star system,
	which components can be viewed from the
	\Aviz{cgi-bin/VizieR-6?-source=1239\&-corr=PK=CCDM\&CCDM==14396-6050}
	{link to the double and multiple stars (CCDM)}
	that appears in the detailed page.
\end{enumerate}

The quantitative monthly usage of \VizieR\ is presently
(October 1999) about 40,000 external requests
from 2700 different nodes;
mirror copies were installed recently in the
\A{http://adc.gsfc.nasa.gov/vizier/}{US} and in
\A{http://z13.mtk.nao.ac.jp/vizier/}{Japan}
in order to overcome the transcontinental network congestions.
%1,000 different nodes effectively
%submitted queries to \VizieR\
%during the first
%3 months of the new installation (June to August 1997);
%among all queries, about 40\% of the hits concern the
%recent results of the \HIP\ and \TYC\ missions.

%\begin{figure}
%\vbox to 5cm { }
%\caption{\label{fig:ftp}Evolution of the usage of the CDS FTP downloading}
%\end{figure}

\subSection{asu}{The ASU protocol}
%From the user's point of view,
The uniform access to all  catalogues  is based on the so-called
\Aviz{doc/asu.html}{ASU} (Astronomical Standardized URL)
protocol resulting from discussions between several institutes
(CDS, ESO, CADC, Vilspa, OAT).
The basic concept of ASU
is a standardized way of specifying queries to remote catalogues
in terms of HTTP requests:
the target catalogue is specified by a \quad
{\tt-source=}{\em catalog\_designation}
parameter,
the target sky position by a \quad
{\tt-c=}{\em name\_or\_position}{\tt,rm=}{\em ra\-dius\_in\_arcmin} parameter,
the output format by {\tt-mime=}{\em type},
and general constraints on parameters by
\quad{\sl column\_name}{\tt=}{\em constraint}.
It should be noticed that the representation of a target by the name of
an astronomical object (typically a star or galaxy name, \eg {\em 3C 273})
implies the usage of a {\em name server} converting a target name
into a position in the sky,
which is typically achieved by a call to {\sc Simbad}.

%All our observations were short direct exposures with
%\htmladdnormallinkfoot{CCD's}{http://www.noao.edu/}.

\subSection{xml}{The XML Interface}

The output of a query to \VizieR\ %for a browsing application
as presented in \secref{netscape} can hardly be used
by an independent  application for further data processing,
%a typical example of a very useful application is represented
such as the \A{http://aladin.u-strasbg.fr/}{{\sc Aladin}}
visualisation tool (Bonnarel et al., \cite{Aladin})
which allows to superimpose the catalogued sources on top of
actual image of the sky:
the application requires an accurate interpretation
of the catalogued output in terms of celestial positions in order
to find out the exact location of each source. This means that
{\sc Aladin} has to figure out not only which are the columns
representing the celestial coordinates, but also accurate definitions
of the system used to express the coordinates, their accuracy, etc\dots ---
in other words the {\em metadata} about  the celestial coordinates.

XML {\em(e{\bf X}tensible {\bf M}arkup {\bf L}anguage)} is an emerging
standard which allows to embed markup {\em ``tags''} within a document;
the key advantages of this language are that
the {\em same document}
can either be parsed by simple-minded programs (XML uses
hierarchical structuring),
or can be displayed in the new generation of browsers (via an
XSL style sheet which maps the markup {\em ``tags''} into typographical
specifications). This language presents other potential interests,
especially regarding interoperability issues facilitated by the emergence
of generic tools able to process XML documents.

The XML layout of astronomical tables was discussed extensively
with interested collaborators, and the agreed definitions were presented
at a recent ADASS meeting (Ochsenbein et al. \cite{adass99}).
The output of \VizieR\ is readily available in this
\A{http://vizier.u-strasbg.fr/cgi-bin/asu-xml}{format}, currently used
by the Aladin image applet; it is
hoped that it will facilitate the usage of the astronomical
data in new contexts.

\subSection{dm}{Current Developments}
With the large set of homogenized catalogues, \VizieR\
plays a central role in a {\em  data-mining project} currently
in development as a collaboration of ESO and CDS, in two main directions:
(i) make use of the \VizieR\ large set of described columns (over
70,000 currently) to build up new methods for locating the
catalogues which are the best suited to a particular research topic;
and (ii) develop automatized cross-correlation tools %between data sets
which can take into account the largest possible set of meaningful parameters
%to make the comparisons
(Ortiz et al, \cite{portiz}).

\Section{conclus}{Conclusions}
\VizieR\ is an illustration of the benefits resulting from
an homogeneous documentation of the existing astronomical catalogues,
facilitating the transformation of a set of heterogeneous data into
a fully interactive database, furthermore able
to interact with remote services.
%adressing also remote services.
The interoperability issues between the databases,
in astronomy and problably in connected disciplines,
will most likely be among the key developments necessary
to allow the scientists to make use of the existing high-quality data
whithout the prerequisite of being familiar with the data.

\begin{acknowledgements}

The long-term exchanges of data %especially with NASA,
have been fundamental for these developments; more specifically,
we wish to thank
Jaylee Mead, Nancy G. Roman, Wayne H. Warren and Gail Schneider at NASA/ADC
for decades of collaborative work, and the present director Cynthia Y. Cheung;
and Olga Dluzhnevskaya at INASAN, the Russian data center.
%Cynthia Cheung.

The support of INSU-CNRS and CNES is acknowledged, as well as the
contribution of ESA-ESIS for the initial developments of \VizieR,
and more specifically Salim Ansari and Isabelle Bourekeb.
The development of \VizieR\ also resulted from fruitful
discussions with Fran\c coi\-se Geno\-va, Michel Cr\'{e}z\'{e} and Daniel Egret;
the enthusiasm of James Lequeux and its implication for the
emergence of electronic tables in the {\em A\&A} publication
had a large impact on the accessiblity of the astronomical data.

We are also grateful to those who contributed in the more tedious,
although critical, part of data standardisation:
Simona Mei, Joseph Florsch and Patricio Ortiz
at CDS; Gail Schneider and collaborators at NASA/ADC;
Koichi Nakajima at ADAC/Japan;
Veta S.Avedisova and collaborators at INASAN;
and we would like to thank also the authors who %, in their vast majority,
participated in the elaboration of the documentation about their data,
and answered patiently to all our questions.

\end{acknowledgements}

\listofobjects
\end{document}